\documentclass[twocolumn,aps,prl,superscriptaddress]{revtex4}
\usepackage[latin9]{inputenc}
\usepackage{amsmath}
\usepackage{amssymb}
\usepackage{bm}
\usepackage{graphicx}
\usepackage{xfrac}
\usepackage{appendix}
\usepackage{CJK}
\usepackage[english]{babel}  
\makeatletter
\@ifundefined{textcolor}{}
{%
 \definecolor{BLACK}{gray}{0}
 \definecolor{WHITE}{gray}{1}
 \definecolor{RED}{rgb}{1,0,0}
 \definecolor{GREEN}{rgb}{0,1,0}
 \definecolor{BLUE}{rgb}{0,0,1}
 \definecolor{CYAN}{cmyk}{1,0,0,0}
 \definecolor{MAGENTA}{cmyk}{0,1,0,0}
 \definecolor{YELLOW}{cmyk}{0,0,1,0}
}

\usepackage{epstopdf}
\usepackage{xcolor}
\usepackage{tikz}

\makeatother

\begin{document}
\title{Predicting Large-Chern-Number Phases in a Shaken Optical Dice Lattice }

\author{Shujie Cheng}
\affiliation{Department of Physics, Zhejiang Normal University, Jinhua 321004, China}
\author{Honghao Yin}
\affiliation{Department of Physics, Capital Normal University,
Beijing 100048, China}
\author{Zhanpeng Lu}
\affiliation{Institute of theoretical physics, Shanxi University, Taiyuan, Shanxi 030006, china}
\author{Chaocheng He}
\affiliation{Department of Physics, Zhejiang Normal University, Jinhua 321004, China}
\author{Pei Wang}
\affiliation{Department of Physics, Zhejiang Normal University, Jinhua 321004, China}
\author{Gao Xianlong}
\affiliation{Department of Physics, Zhejiang Normal University, Jinhua 321004, China}
\date{\today}

\begin{abstract}
  With respect to the quantum anomalous Hall effect (QAHE), the detection of topological nontrivial large-Chern-number phases is an intriguing subject. Motivated by recent research on Floquet topological phases, this study proposes a periodic driving protocol to engineer large-Chern-number phases using QAHE. Herein, spinless ultracold fermionic atoms are studied in a two-dimensional optical dice lattice with nearest-neighbor hopping and a $\Lambda$/V-type sublattice potential subjected to a circular driving force. Results suggest that large-Chern-number phases exist with Chern numbers equal to $C=-2$, which is consistent with the edge-state energy spectra.
 \end{abstract}

\maketitle

\section{Introduction}
In recent decades, since the discovery of the quantum Hall effect at low temperatures with strong magnetic fields \cite{Klitzing},
many studies have been conducted on the topological features of condensed matter \cite{TKNN,Haldane_model,
 TI_1,TI_2,TI_4,TI_6,TI_8,TI_9,bulk-edge}.
 Quantum anomalous Hall effect (QAHE) is a branch in this sustainable field\cite{TI_8}, and Chern
 insulators are one of the many insulators used in QAHE. Different from other topological insulators with
 time inversion symmetry \cite{TI_4}, Chern insulators have topological features characterized by the Chern number ($C$), also known as the
 topological invariant, which was first proposed by Thouless-Kohmoto-Nightingale-den Nijs (TKNN) \cite{TKNN}. Moreover, $C\neq 0$ ($C=0$)
 corresponds to the topological non-trivial (trivial) phase.

A paradigmatic Chern insulator system is the Haldane model \cite{Haldane_model}. Motivated by this model, some other lattice systems \cite{long_distance_2,long_distance_3,Checkerboard,Kagome_1,Kagome_2,Kagome_3,Kagome_4,Lieb_1,Lieb_3,Lieb_4,
 dice_model_1,dice_model_2,dice_model_3} are also predicted to contain topological non-trivial phases. Of these, large-Chern-number
 phases with QAHE have attracted widespread attention. In particular, some studies have considered long-range
 tunneling \cite{long_distance_2,long_distance_3} or complex magnetic
 flux \cite{dice_model_1,dice_model_2,dice_model_3} to obtain
topological phases with large Chern numbers. Further, large-Chern-number phases are theoretically predicted to
 appear in multilayer crystal structures \cite{multi_1,multi_3,multi_4}. In the field of interface transport, such intriguing phases are
 expected to enhance the performance of certain devices by reducing the channel resistance \cite{device_1,device_2}. However, owing to the limitations
of lattice structure, observing large-Chern-number phases in reality remains impossible. Even in topological nontrivial materials, the Chern numbers are mostly detected for $C=\pm 1$ \cite{observe_3,observe_4,observe_5}. Rarely,
 in the context of the photonic crystals \cite{photonic}, Chern numbers are measured up to $C=4$ by introducing an external magnetic field.

Recently, the periodic driving protocol has been developed to generate the Floquet topological nontrivial phases with QAHE. Its flexibility and
versatility realize the possible designing of Floquet topological bands on demand; the scenario of Floquet topological states has been studied in many systems \cite{shaking,Floquet_1,Floquet_2,Floquet_3,Floquet_4,Floquet_5,Floquet_6,Floquet_7,Floquet_8,
 Floquet_9,Floquet_10,Floquet_11,Floquet_theory_4,Floquet_theory_5,Floquet_large_CN,Floquet_large_CN_1}. Moreover, the Floquet engineering is also recoganized as an effective approach  to realize the topological phases with large topological invariants
 \cite{Floquet_large_CN,Floquet_large_CN_1}. This study aims to exploit a
 periodic driving protocol applied to a two-dimensional (2D) dice lattice \cite{dice_lattice_2,dice_lattice_3,dice_lattice_4,dice_lattice_5,
 dice_lattice_6,dice_lattice_7,dice_lattice_8,dice_lattice_9,dice_model_1,dice_model_2,dice_model_3} to realize large-Chern-number phases
 with QAHE. To this end, spinless ultracold fermionic atoms were studied in a 2D optical dice lattice with nearest-neighbor hopping and a
 $\Lambda$/V-type sublattice potential subjected to a circular driving force $\bm{F}(t)=F\left[\cos(\omega {t})\bm{e_x}-\sin(\omega {t})\bm{e_y}\right]$ \cite{shaking}.
 Here, $F$ and $\omega$ are the amplitude and frequency of the driving force, respectively. The proposed protocol is based on two
 considerations: lattice geometry, which can be created using laser beams \cite{stand_wave_3}, and
 shaking \cite{shaking}, like the Haldane model \cite{Haldane_model}. We hope this system can be realised using similar experimental settings.
 After calculating the effective Hamiltonian, large-Chern-number phases are identified that contain $C=-2$; these results are consistent with the edge-state energy spectra.

 The remaining portions of this article is organized as follows:~Sec.~\ref{S2}, describes the periodically driven dice model, which can be realized using the ultracold atoms trapped in a circularly shaken 2D dice optical lattice, presenting the effective Hamiltonian derived from the Floquet theorem in the process; Sec.~\ref{S3} maps the effective Hamiltonian into the SU(3) system and numerically obtains the Chern-number phases; the bulk-edge correspondence was also analyzed; finally, a brief summary is presented in Sec.~\ref{S4}.

\section{\label{S2}PERIODICALLY DRIVEN DICE MODEL}
 \begin{figure}[t]
  \centering
  \includegraphics[width=0.45\textwidth]{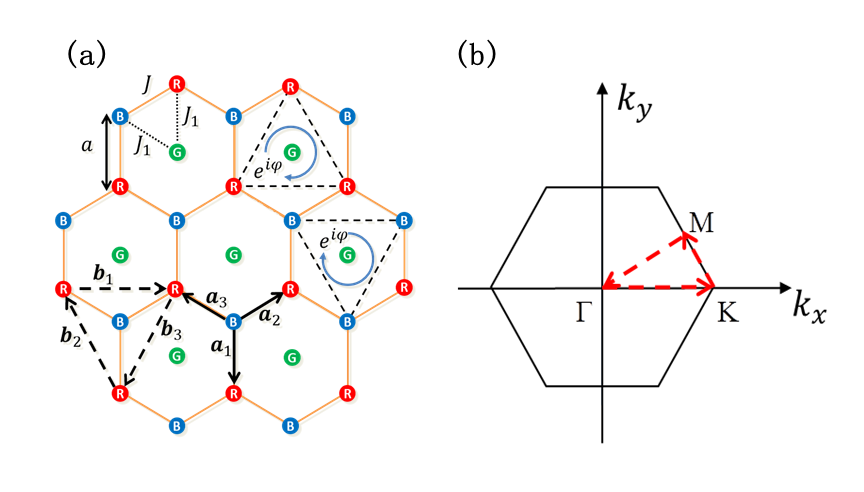}
  \caption{(Color Online) (a) Schematic of the dice lattice with sublattice R (red), B (blue), and G (green). The Hamiltonian $\hat{H}_{\rm tun}$ possesses real tunneling matrix elements $J$ between neighboring R and B sites and $J_1$ between neighboring G and R/B sites; the circular arrows show the tunneling between the same R/B sublattice sites with tunneling phase $e^{i\varphi}$ ($\varphi=\frac{\pi}{2}$), which is contained in the effective Hamiltonian $\hat{\mathcal{H}}_{\rm eff}$. The vectors $\bm{a}_{s}$ ($s=1,2,3$) connect the nearest neighbor sites, whereas vectors $\bm{b}_{s}$ ($s=1,2,3$) connect the next nearest-neighbor R/B sites. (b) The first Brillouin zone of the dice lattice. $\Gamma$, K and M are high symmetry points, connected by three dashed red arrows.
  }\label{f1}
 \end{figure}

Spinless ultracold fermionic atoms trapped in a shaken 2D dice optical lattice were considered, as depicted in Fig.~\ref{f1}(a).
Three interpenetrating triangle sublattices were present: R, B, and G.
 This lattice geometry is similar to the Haldane model and can be formed using three retro-reflected laser beams \cite{stand_wave_3}.
 Through tight-binding approximation, the single particle Hamiltonian of this shaken lattice system can be expressed as
 \begin{equation}
 \hat{H}=\hat{H}_{\rm tun}+\hat{H}_{\rm pot}+\hat{H}_{\rm dri}.
 \end{equation}

 The first term $\hat{H}_{\rm tun}$ indicates the tunneling kinetics and can be expressed as
 \begin{equation}
 \begin{aligned}
 \hat{H}_{\rm tun}&=\sum_{\langle R_j,B_{j'}\rangle}J\left(\hat{c}^\dag_{R_j}\hat{c}_{B_{j'}}+H.c.\right)\\
 &+\sum_{\langle G_{j},R_{j'}\rangle}J_1\left(\hat{c}^\dag_{G_j}\hat{c}_{R_{j'}}+H.c.\right)\\
 &+\sum_{\langle B_{j},G_{j'}\rangle}J_1\left(\hat{c}^\dag_{B_j}\hat{c}_{G_{j'}}+H.c.\right),
 \end{aligned}
 \end{equation}
where the sum extending over the nearest-neighbor sites, $R_j$, $B_j$, and $G_j$, denotes the relevant sites of
 sublattice R,~B, and G, respectively, and $j$ denotes the corresponding site index; $J$ denotes the tunneling parameter
 between one R site and one B site; $J_1$ denotes the tunneling parameter between one R/B site and one G site; and
 $\hat{c}_{R_j}, \hat{c}_{B_j}$, $\hat{c}_{G_j}$ denote the corresponding fermionic annihilation operators.

 The second term $\hat{H}_{\rm pot}$ describes a special on-site potential for atoms populating various sublattices:
 \begin{equation}
 \hat{H}_{\rm pot}=\epsilon_{R}\sum_{R_j}\hat{c}^\dag_{ R_j}\hat{c}_{R_{j}}+\epsilon_{B}\sum_{B_j}\hat{c}^\dag_{B_j}\hat{c}_{B_{j}}
 +\epsilon_{G}\sum_{G_j}\hat{c}^\dag_{G_j}\hat{c}_{G_{j}},
 \end{equation}
where R and B sublattices have the same on-site potential $\epsilon_{R}=\epsilon_{B}\equiv\gamma_{1}\Delta$ and the G sublattice has the on-site potential of $\epsilon_{G}\equiv\gamma_{2}\Delta$, where $\Delta$ denotes the strength of the potential and $\gamma_{1}$ and $\gamma_{2}$ reflects how fast the tunable potential energy changes. If $\gamma_{1}$ and $\gamma_2$ are both positive and $\gamma_{1}<\gamma_{2}$, the potential
 presents a $\Lambda$- (V-)-type structure when $\Delta>0$ ($\Delta<0$).
 The $\Lambda$- (V-) structure also appears in other cases where $\gamma_{1}$ and $\gamma_{2}$ are both completely or partially negative.
 Such on-site potential is different from that presented in previous studies on
 dice models \cite{dice_model_1,dice_model_2,dice_model_3}; it can be realized by tuning the single-beam lattice depths.

The third term $\hat{H}_{\rm dri}$ denotes the contribution of the circular periodic driving force in terms of the time-dependent on-site potential; This third term can be expressed as
 \begin{equation}
 \hat{H}_{\rm dri}=\sum_{\alpha_{j}}V(\bm{r}_{\alpha_{j}},t)\hat{n}_{\alpha_{j}},
 \end{equation}
 where the sum runs over all lattice sites; $\bm{r}_{\alpha_{j}}$ denotes the lattice-site coordinates; $\alpha~\epsilon$~$\{$R,G,B$\}$ denotes
sublattice type; $\hat{n}_{\alpha_{j}}=\hat{c}^\dag_{\alpha_{j}}\hat{c}_{\alpha_{j}}$
 are the site number operators; and $V(\bm{r}_{\alpha_{j}},t)=-\bm{r}_{\alpha_{j}}\cdot\bm{F}(t)$.

Herein, the case of strong driving is explored, wherein the amplitude $F$ is scaled according to the driving frequency; i.e., $Fa \sim \hbar\omega$.
 Therefore, performing a gauge transformation before employing high-frequency
 approximation \cite{Floquet_theory_4,Floquet_theory_5} is necessary. The gauge-dependent time-periodic unitary operator can be expressed as
 \begin{equation}
 \hat{U}(t)=\exp\left(-\frac{i}{\hbar}\sum_{\alpha_{j}}
 \int^{t}_{0}V(\bm{r}_{\alpha_{j}},t)dt\cdot\hat{n}_{\alpha_{j}}\right).
 \end{equation}
 Next, the transformed time-periodic Hamiltonian is determined (see the details provided in Appendix A):
 \begin{equation}
 \begin{aligned}
 \hat{H}_{\rm tra}&=\hat{U}^\dag(t)\hat{H}\hat{U}(t)-i\hbar\hat{U}^\dag(t)\dot{\hat{U}}(t)\\
 &=\hat{U}^\dag(t)\hat{H}_{\rm tun}\hat{U}(t)+\hat{H}_{\rm pot}.
 \end{aligned}
 \end{equation}

 The important features of the above time-dependent Hamiltonian $\hat{H}_{\rm tra}$ can be acquired from an effective time-independent Hamiltonian, which can be obtained by
 expanding the Hamiltonian $\hat{H}_{\rm tra}$ as follows:
 \begin{equation}
 \hat{H}_{\rm tra}=\sum_{m=-\infty}^{\infty}\hat{\mathcal{H}}_{m} e^{i m \omega t}+\hat{H}_{\rm pot},
 \end{equation}
 where $\hat{\mathcal{H}}_m$ denotes the Fourier components, modified by the Bessel
 functions $\mathcal{J}_{m}(\beta)$ (see the derivation in Appendix B). Then, the effective Hamiltonian with high-frequency
 approximation can be obtained by truncating the high-frequency expansion into finite orders as follows:
 \begin{equation}
 \hat{\mathcal{H}}_{\rm eff}=\hat{\mathcal{H}}_{0}+ \sum^{\infty}_{m=1} \frac{1}{m\hbar\omega}\left[\hat{\mathcal{H}}_{m}, \hat{\mathcal{H}}_{-m}\right]+\hat{H}_{\rm pot}.
 \end{equation}

 \noindent The high-order terms of $1/\omega$ can be neglected in high-frequency driving
 cases \cite{Floquet_10,Floquet_theory_4,Floquet_theory_5}.

\begin{figure}[t]
  \centering
  \includegraphics[width=0.45\textwidth,height=0.25\textheight]{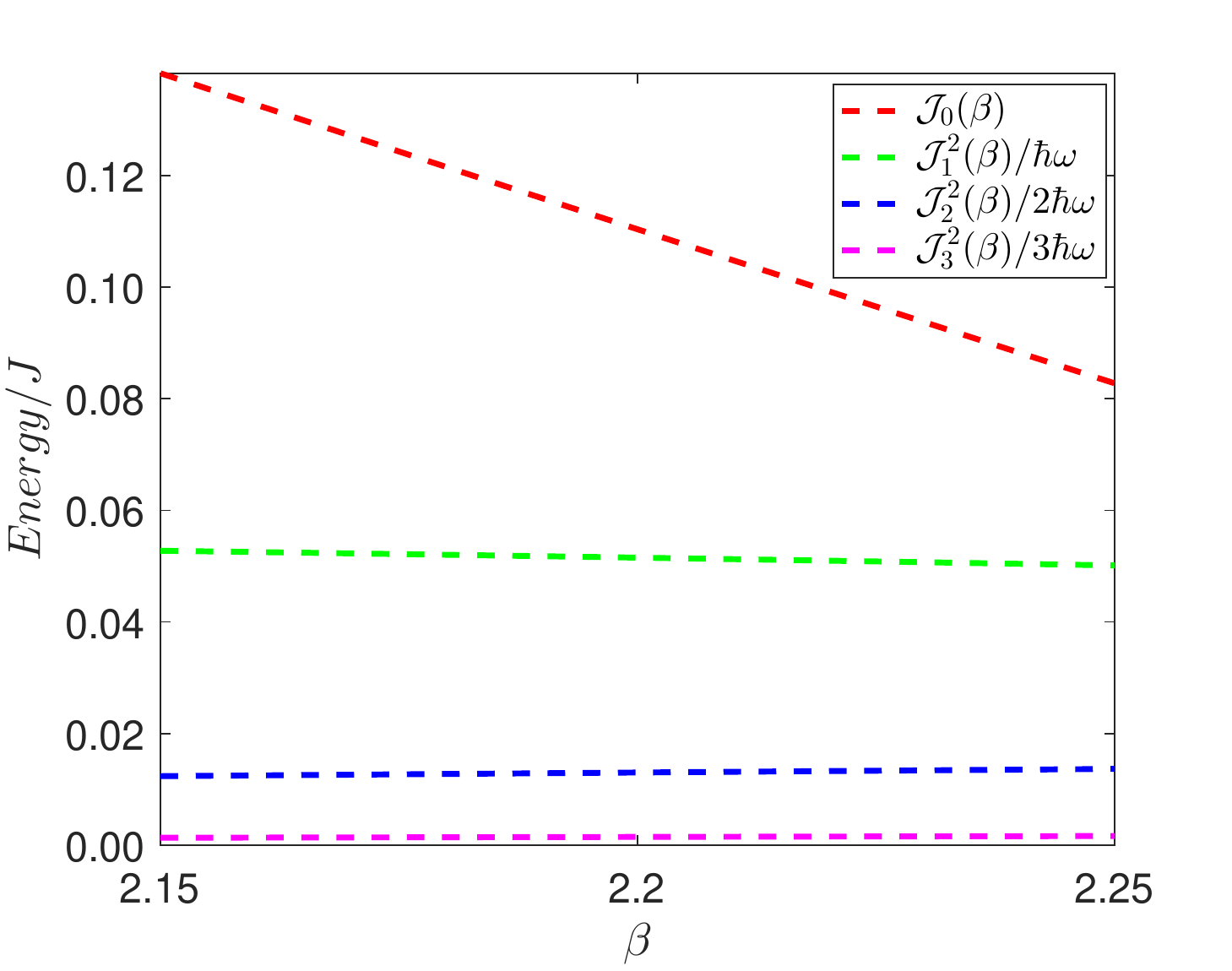}\\
  \caption{(Color Online) Zero-order Bessel function (dashed red line), as well as the square of the first-, second-, and
  third-order Bessel function divided by the product of its' order number and $\hbar\omega$ (dashed green, blue, and magenta lines, respectively). }\label{f2}
 \end{figure}

 In the theoretical analysis, $\hbar\omega=6J$ and $\beta=2.2$ are considered as an example; the terms in $\hat{\mathcal{H}}_{m}$
 containing $\mathcal{J}_{m}(\beta) (m\geq {3})$ are negligible (see Fig.~\ref{f2}). Therefore, the sum in $\hat{\mathcal{H}}_{\rm eff}$ is
 truncated into three terms. Thus,
 \begin{equation}
 \begin{aligned}
 \hat{\mathcal{H}}_{\rm eff}&=\hat{\mathcal{H}}_{0}+\sum_{m=1,2}\frac{1}{m\hbar\omega}\left[\hat{\mathcal{H}}_{m},
 \hat{\mathcal{H}}_{-m}\right]+\hat{H}_{\rm pot}\\
 =&\left[\sum_{\langle R_{j},B_{j'}\rangle}J_{rb}\hat{c}^\dag_{R_j}\hat{c}_{B_{j'}}
 +\sum_{\langle B_{j}, G_{j'}\rangle}J_{bg}\hat{c}^\dag_{B_j}\hat{c}_{G_{j'}}\right.\\
 +&\left.\sum_{\langle G_{j},R_{j'}\rangle}J_{gr}\hat{c}^\dag_{G_j}\hat{c}_{R_{j'}}
 +\sum_{\ll R_{j},R_{j'}\gg}J_{rr}\hat{c}^\dag_{R_j}\hat{c}_{R_{j'}}\right.\\
 +&\left.\sum_{\ll B_{j},B_{j'}\gg}J_{bb}\hat{c}^\dag_{B_j}\hat{c}_{B_{j'}}+H.c.\right]+\epsilon_{R}\sum_{ R_j}\hat{c}^\dag_{ R_j}\hat{c}_{R_{j}}\\
 +&\epsilon_{B}\sum_{B_j}\hat{c}^\dag_{B_j}\hat{c}_{B_{j}}
 +\epsilon_{G}\sum_{G_j}\hat{c}^\dag_{G_j}\hat{c}_{G_{j}}\,\,,
 \end{aligned}
 \end{equation}
 where $\ll\cdots\gg$ denotes the next nearest-neighbor relations, and the tunneling parameters are
 \begin{eqnarray}
 \begin{aligned}
 &J_{rb}=J\mathcal{J}_{0}\left(\beta\right),\\
 &J_{bg/gr}=J_{1}\mathcal{J}_{0}\left(\beta\right),\\
 &J_{rr/bb}=\frac{\sqrt{3}}{2}e^{i\sigma\frac{\pi}{2}}\left(\frac{J^2-J_1^2}{\hbar\omega}\right)\left[\mathcal{J}^2_{1}\left(\beta\right)-\frac{1}{2}\mathcal{J}^2_{2}\left(\beta\right)\right],
 \end{aligned}
 \end{eqnarray}
 with $\sigma=1$ for clockwise tunneling and $\sigma=-1$ for counterclockwise tunneling.

 \section{\label{S3}CHERN NUMBER AND EDGE STATE}
 The infinite system was considered such that the translational symmetry is preserved. At $1/3$ or $2/3$ filling~\cite{su31a}, the $\hat{\mathcal{H}}_{\rm eff}$ can be mapped into the SU(3) system, and the general Bloch Hamiltonian \cite{su3_1,su3_3}
 can be expressed as
 \begin{equation}
 \hat{\mathcal{H}}(\mathbf{k})=I(\mathbf{k})+\mathbf{d}(\mathbf{k})\cdot\vec{\lambda},
 \end{equation}
where $\mathbf{k}$ denotes a wave vector; $I(\mathbf{k})$ denotes a scalar;~$\mathbf{d}(\mathbf{k})$ denotes an eight-dimensional real vector; and
 $\vec{\lambda}$
 denotes a vector of Gell--Mann matrices \cite{Gell}. In practice, the scalar $I(\mathbf{k})$ has no effect on the wave function; hence, the Chern number
 can only be determined by the coefficient vectors $\mathbf{d}(\mathbf{k})$. After performing discrete Fourier transformation in the three-component
 basis, $\left(\hat{c}_{\mathbf{k},R},\hat{c}_{\mathbf{k},B},\hat{c}_{\mathbf{k},G}\right)^T$, where
 $\hat{c}_{\mathbf{k},\alpha}=\frac{1}{\sqrt{N}}\sum_{\alpha_j}e^{-i\mathbf{k}\cdot\bm{r}_{\alpha_j}}\hat{c}_{\alpha_j}$
 ($N$ is the unit-cell number), and the components of the vector $\bf{d}(\bf{k})$ can be given as
 \begin{equation}
 \begin{aligned}
 d_1&=J_{rb}\sum_{s}\cos\left(\mathbf{k}\cdot\bm{a}_s\right),~~d_2=J_{rb}\sum_{s}\sin\left(\mathbf{k}\cdot\bm{a}_s\right),\\
 d_4&=d_6=J_{gr/bg}\sum_{s}\cos\left(\mathbf{k}\cdot\bm{a}_s\right),\\
 d_7&=-d_5=J_{gr/bg}\sum_{s}\sin\left(\mathbf{k}\cdot\bm{a}_s\right),\\
 d_3&=-2J_{rr/bb}\sum_{s}\sin(\mathbf{k}\cdot\bm{b}_s),
 ~~d_8=\frac{\gamma_{1}-\gamma_{2}}{\sqrt{3}}\Delta,
 \end{aligned}
 \end{equation}
 where, we have set $a=1$ for convenience and the six vectors $\bm{a}_s$ and $\bm{b}_s$ ($s=1,2,3$), which are shown
 in Fig.~\ref{f1}(a), are displayed as
 \begin{equation}
 \begin{aligned}
 \bm{a}_1&=\binom{0}{-1},\quad\bm{a}_2=\frac{1}{2}\binom{\sqrt{3}}{1},
 \quad\bm{a}_3=\frac{1}{2}\binom{-\sqrt{3}}{1},\\
 \bm{b}_1&=\binom{\sqrt{3}}{0},\quad\bm{b}_2=\frac{1}{2}\binom{-\sqrt{3}}{3},
 \quad\bm{b}_3=-\frac{1}{2}\binom{\sqrt{3}}{3}.
 \end{aligned}
 \end{equation}

\begin{figure}[t]
  \centering
  \includegraphics[width=0.45\textwidth]{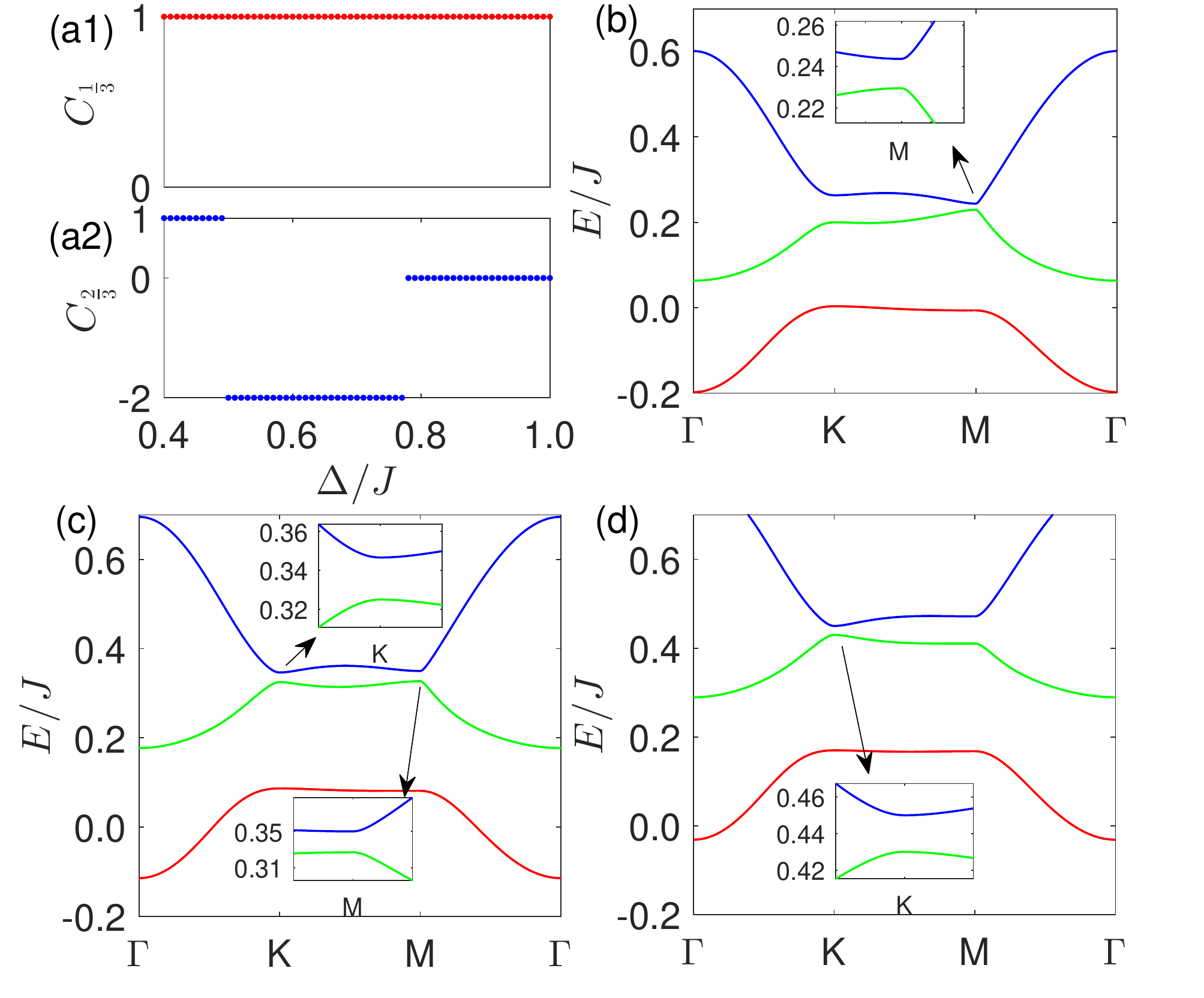}\\
  \caption{(Color Online)
  (a1) Phase diagram of the Chern numbers $C_{\frac{1}{3}}$ for the gapped lowest band ($1/3$ filling).
  The system always stays in the topological nontrivial phase with $C_{\frac{1}{3}}=1$.
  (a2) The corresponding phase diagram of the Chern numbers $C_{\frac{2}{3}}$ for the gapped lower two bands ($2/3$ filling). The system remains in the large-Chern-number phase with $C_{\frac{2}{3}}=-2$.
 (b)--(d) Gapped dispersions along the high-symmetric path $\Gamma$ - K - M - $\Gamma$ (see Fig. 1(b)) with parameters
 $\Delta=0.4 J$, $\Delta=0.65 J$, and $\Delta=0.9 J$, respectively. The other parameter used is $J_1=0.5 J$.
 The insets are enlargements of the dispersions near the high-symmetric points indicated by the arrows.}\label{f3}
\end{figure}

Here, the system is shown to contain gapped phases with Chern numbers larger than one; these
 topological states are supported by the edge-state energy spectra. For any given occupied band, the Chern number is
 defined as \cite{dice_model_3,define_CN_1,define_CN_2}
 \begin{equation}
 C_{n}=\frac{1}{2 \pi} \oint_{\partial{BZ}} \mathbf{A}_{n}(\mathbf{k}) \cdot d \mathbf{k},
\end{equation}
where $\partial{BZ}$ denotes the boundary of the first Brillouin zone; $n$ denotes the band index; $\mathbf{A}_{n}$ denotes the Berry connection
 with $\mathbf{A}_{n}=-i\left\langle\psi_{n}(\mathbf{k})\left|\nabla_{\mathbf{k}}\right| \psi_{n}(\mathbf{k})\right\rangle$; and $\left|\psi_{n}(\mathbf{k})\right\rangle$
 denotes the corresponding eigenvector of $\hat{\mathcal{H}}(\mathbf{k})$.
A slow-regulated $\Lambda$- or V-type potential with $\gamma_{1}=\frac{1}{3}$ and $\gamma_{2}=\frac{1}{2}$ is considered in the rest of the numerical calculations~\cite{be65}. The Chern numbers are calculated according to Eq.~(14). $C_{\frac{1}{3}}$ is used to denote
 the Chern number of $1/3$ filling and $C_{\frac{2}{3}}$ for $2/3$ filling.

 Fig.~\ref{f3}(a1) and Fig.~\ref{f3}(a2) are the phase diagrams that exhibit variations in the Chern number as a function of the parameter $\Delta$.
 In Fig.~\ref{f3}(a1), $C_{\frac{1}{3}}$ maintains a plateau within the interval of $\Delta$~(shown by the dotted red line). Therefore, the
 system is always in the topological nontrivial phase with $C_{\frac{1}{3}}=1$. In Fig.~\ref{f3}(a2), useful phases appear when the system is at $2/3$ filling (shown by the dotted blue lines). For small values of $\Delta$, the system is in the topological nontrivial phase with
 $C_{\frac{2}{3}}=1$. When $\Delta$ increases, the system undergoes a phase transition, entering a large-Chern-number phase with
 $C_{\frac{2}{3}}=-2$. With the sustainable growth of $\Delta$, the system becomes trivial with $C_{\frac{2}{3}}=0$. To illustrate that these topological phases are gapped, we plot the dispersion relations at three chosen parameters ($\Delta=0.4 J,~0.65 J$, and $ 0.9J $) presented in Fig.~\ref{f3}(b), Fig.~\ref{f3}(c), and Fig.~\ref{f3}(d), respectively. In each diagram,
 the insets show the local amplification of the dispersion near the high-symmetric points \cite{Kpoints}. As can be observed in the diagrams,
 no bands touch each other.
 \begin{figure}[t]
  \centering
  \includegraphics[width=0.45\textwidth]{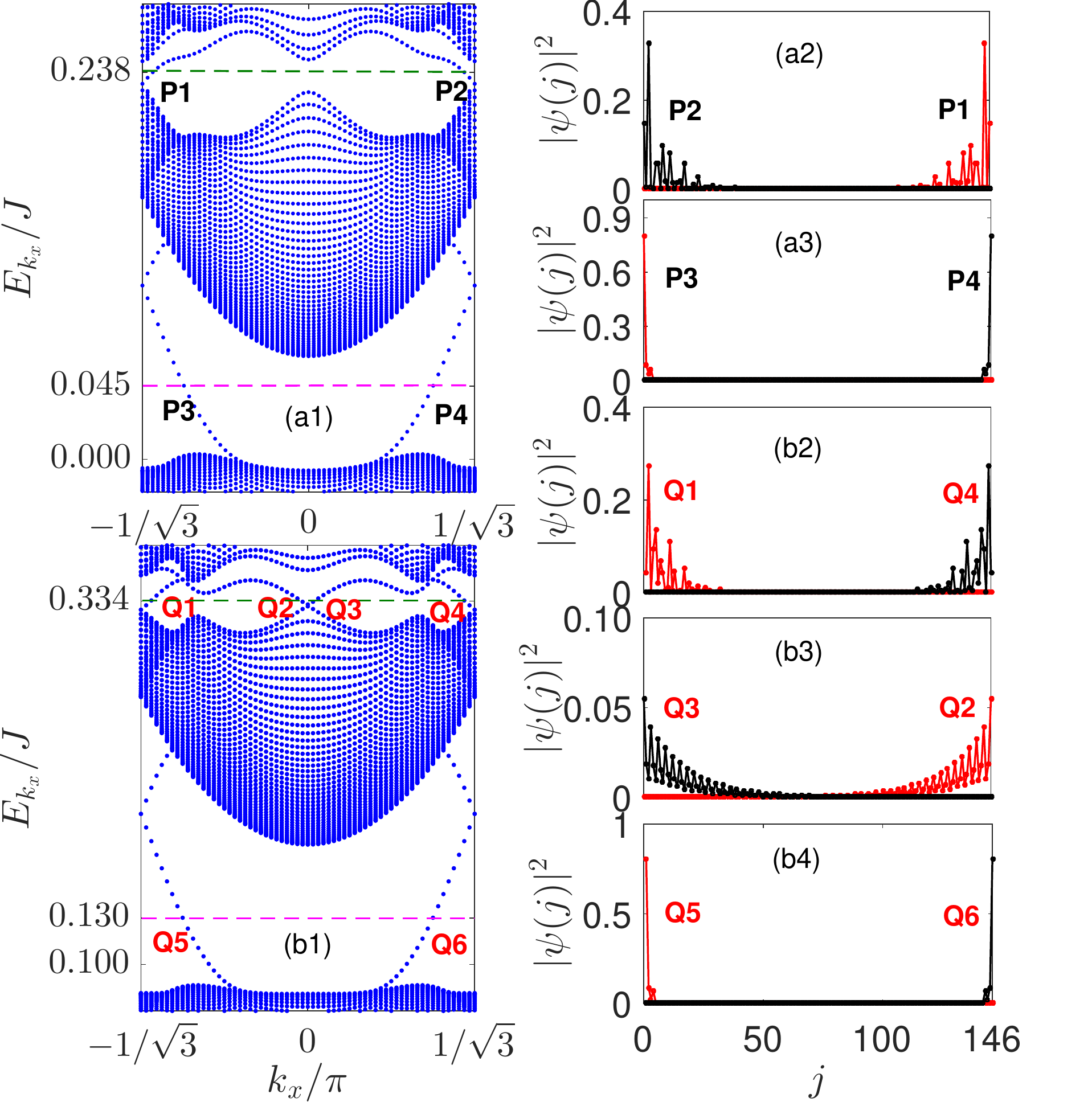}\\
  \caption{(Color Online) Two edge-state spectra under the periodic boundary condition in the $x$ direction with the open boundary
 condition in the $y$ direction. (a1) $\Delta=0.4 J $. For $2/3$ filling, the Fermi energy is chosen as $0.238 J$ (dashed green line).
In this case, a pair of edge modes exist, labeled as P1 and P2. For $1/3$ filling, the Fermi energy is chosen as $0.045 J$ (dashed magenta line). Another pair of edge
 modes, labeled P3 and P4, also exist. (b1) $\Delta=0.65 J$. For $2/3$ filling, the Fermi energy is chosen as $0.334 J$ (dashed green line). In this case, two pairs of edge modes,
 labeled Q1 and Q4 and  Q2 and Q3 exist. For $1/3$ filling, the Fermi energy is chosen as $0.130 J$ (dashed magenta line). Here, only one pair of edge modes, labeled Q5 and Q6 exists. The spatial density distributions of these paired edge modes are plotted in (a2)--(a3) and (b2)--(b4), respectively. Results suggest that the edge states with opposite momentums are localized on different system boundaries, presenting the chiral symmetry. A total of 146 lattice sites are considered in the numerical calculations. }\label{f4}
 \end{figure}

 Here, the bulk-edge correspondence of the topological systems  is discussed \cite{bulk-edge,dice_model_1,dice_model_2,dice_model_3}.
In Hermitian systems, the Chern number can precisely count the number of topological edge states
 observed in the edge-state spectrum. The chiral edge states were studied by
 considering a cylindrical geometry with periodic boundary conditions in the $x$ direction and open boundary conditions in the
 $y$ direction. The Hamiltonian can be acquired via partial Fourier transformation and is parameterized by the
 good quantum number $k_x$. By choosing $\Delta=0.4 J$ ($\Delta=0.65 J$), the edge-state spectra are plotted as a function
 of $k_x$, as shown in Fig.~\ref{f4}(a) (Fig.~\ref{f4}(b)). Intuitively, when $\Delta=0.4 J$, the edge-state spectrum
 intersects at $k_x=\pm \frac{\pi}{\sqrt{3}}$ within both bulk gaps, implying a pair of chiral edge states
 regardless of the filling, corresponding to $C_{\frac{1}{3}}=C_{\frac{2}{3}}=1$. Conversely, when $\Delta=0.65 J$, within the upper bulk gap, the edge-state spectrum intersects in two different manners, occurring at $k_x=\pm \frac{\pi}{\sqrt{3}}$ and $k_x=0$.
 This indicates two pairs of chiral edge states at the $2/3$ filling corresponding to $C_{\frac{2}{3}}=-2$.
 At $1/3$ filling, only one pair of chiral edge states is observed, corresponding to $C_{\frac{1}{3}}=1$. In Fig.~\ref{f4}(a) and Fig.~\ref{f4}(b), the dashed green and magenta lines represent the chosen Fermi energies. P1--P4 and Q1--Q6 are the corresponding edge modes. To characterize the edge state localization, the spatial density distributions of the corresponding edge modes were plotted in Fig.~\ref{f4}(a2--a3) and Fig.~\ref{f4}(b2--b4), respectively. Results suggest that edge states with opposite $k_x$ are localized on different system boundaries, presenting the chiral symmetry.

 \section{\label{S4}SUMMARY}
 Herein, a periodic driving protocol was proposed to engineer large-Chern-number phases with QAHE in a 2D
 periodically shaken optical dice model. Using the Floquet method, phase diagrams with $1/3$ filling and $2/3$ filling were obtained.
The analytical results suggest that large-Chern-number phases exist with $C=-2$ at $2/3$ filling, which is consistent with the edge-state spectra.
This theoretical model can be implemented using similar experimental settings of realizing the Haldane model \cite{shaking}.
Moreover, the proposed protocol is beneficial for identifying large-Chern-number phases in other Hermitian and non-Hermitian systems.

\section{acknowledgments}

GX and CJ acknowledge support from NSFC under
Grants No. 11835011 and No. 11774316.

 \begin{appendix}
 \section{ Derivation of Equation (6)}
 The transformed Hamiltonian $\hat{H}_{\rm tra}$ in Eq.~(6) can be expanded as
 \begin{equation}
 \begin{aligned}
 \hat{H}_{\rm tra}&=\hat{U}^\dag(t)\hat{H}\hat{U}(t)-i\hbar\hat{U}^\dag(t)\dot{\hat{U}}(t)\\
 &=\hat{U}^\dag(t)\hat{H}_{\rm tun}\hat{U}(t)+\hat{U}^\dag(t)\hat{H}_{\rm pot}\hat{U}(t)\\
 &=\sum_{\langle R_j,B_{j'}\rangle}J\left(\hat{U}^\dag(t)\hat{c}^\dag_{ R_j}\hat{U}(t)\hat{U}^\dag(t)\hat{c}_{ B_{j'}}\hat{U}(t)+H.c.\right)\\
 &+\sum_{\langle G_{j}, R_{j'}\rangle}J_1\left(\hat{U}^\dag(t)\hat{c}^\dag_{ G_j}\hat{U}(t)\hat{U}^\dag(t)\hat{c}_{ R_{j'}}\hat{U}(t)+H.c.\right)\\
 &+\sum_{\langle B_{j}, G_{j'}\rangle}J_1\left(\hat{U}^\dag(t)\hat{c}^\dag_{ B_j}\hat{U}(t)\hat{U}^\dag(t)\hat{c}_{ G_{j'}}\hat{U}(t)+H.c.\right)\\
 &+\epsilon_{R}\sum_{R_{j}}\hat{U}^\dag(t) \hat{n}_{R_{j}}\hat{U}(t)+\epsilon_{B}\sum_{B_{j}}\hat{U}^\dag(t) \hat{n}_{B_{j}}\hat{U}(t)\\
 &+\epsilon_{G}\sum_{G_{j}}\hat{U}^\dag(t)\hat{n}_{G_{j}}\hat{U}(t),
 \end{aligned}
 \end{equation}
 where $\hat{U}(t)=\exp \left(-\frac{i}{\hbar}\sum_{\alpha_{j}}\chi_{\alpha_{j}}(t)\cdot \hat{n}_{\alpha_{j}}\right)$ with $\chi_{\alpha_{j}}(t)=\int^{t}_{0}V(\mathbf{r}_{\alpha_{j}},t)dt$.

Using the expansions presented above \cite{Floquet_theory_5},
 \begin{equation}
 \begin{aligned}
 e^{i\hat{X}}\hat{Y}e^{-i\hat{X}}&=\hat{Y}+i[\hat{X},\hat{Y}]-\frac{1}{2}[\hat{X},[\hat{X},\hat{Y}]]\\
 &-\frac{i}{6}[\hat{X},[\hat{X},[\hat{X},\hat{Y}]]]\ldots
 \end{aligned}
 \end{equation}
 and considering the relation $\left[\hat{n}_{\alpha_{j}},\hat{n}_{\alpha'_{j'}}\right]=0$,  $\hat{H}_{tr}$ was finally obtained, which can be
 expressed as
 \begin{equation}
 \begin{aligned}
 \hat{H}_{\rm tra}&=\sum_{\langle R_j,B_{j'}\rangle}J\left(e^{-i\frac{Fa}{\hbar\omega}\sin(\omega t+\theta^{R_{j}}_{B_{j'}})}\hat{c}^\dag_{R_j}\hat{c}_{B_{j'}}+H.c.\right)\\
 &+\sum_{\langle G_{j},R_{j'}\rangle}J_1\left(e^{-i\frac{Fa}{\hbar\omega}\sin(\omega t+\theta^{G_{j}}_{R_{j'}})}\hat{c}^\dag_{G_j}\hat{c}_{R_{j'}}+H.c.\right)\\
 &+\sum_{\langle B_{j},G_{j'}\rangle}J_1\left(e^{-i\frac{Fa}{\hbar\omega}\sin(\omega t+\theta^{B_{j}}_{G_{j'}})}\hat{c}^\dag_{B_j}\hat{c}_{G_{j'}}+H.c.\right)\\
 &+\hat{H}_{\rm pot},
 \end{aligned}
 \end{equation}
 where the angle $\theta^{\alpha_{j}}_{\alpha'_{j'}}$ is defined by the direction of the vector pointing from site $\alpha'_{j'}$ to its neighbor $\alpha_{j}$,
 \begin{equation}
 \bm{r}_{\alpha_{j}}-\bm{r}_{\alpha'_{j'}}=a\left[\cos\left(\theta^{\alpha_{j}}_{\alpha'_{j'}}\right)\bm{e}_{x}+\sin\left(\theta^{\alpha_{j}}_{\alpha'_{j'}}\right)\bm{e}_{y}\right].
 \end{equation}

 \section{Derivation of $\hat{\mathcal{H}}_{m}$}
In this appendix, Eq. (7) is derived.
 For convenience, we set
 \begin{equation}
 \begin{aligned}
 h&=\exp\left[-i\frac{Fa}{\hbar\omega}\sin(\omega t+\theta^{\alpha_{j}}_{\alpha'_{j'}})\right]\\
 &=\exp\left[\frac{Fa}{\hbar\omega}\cdot\frac{e^{-i\sin(\omega t+\theta^{\alpha_{j}}_{\alpha'_{j'}})}-e^{i\sin(\omega t+\theta^{\alpha_{j}}_{\alpha'_{j'}})}}{2}\right].
 \end{aligned}
 \end{equation}

 By replacing $\frac{Fa}{\hbar\omega}$ with $\beta$ and using the Bessel function
 \begin{equation}
 \exp\left[\xi\frac{x-x^{-1}}{2}\right]=\sum^{\infty}_{\ell=-\infty}\mathcal{J}_{\ell}(\xi)x^{\ell},
 \end{equation}
 we can obtain
 \begin{equation}
 \begin{aligned}
 h&=\sum^{\infty}_{\ell=-\infty}\mathcal{J}_{\ell}(\beta)e^{-i\ell(\omega t+\theta^{\alpha_{j}}_{\alpha'_{j'}})},\\
 h^{*}&=\sum^{\infty}_{\ell=-\infty}\mathcal{J}_{\ell}(\beta)e^{i\ell(\omega t+\theta^{\alpha_{j}}_{\alpha'_{j'}})}.
 \end{aligned}
 \end{equation}

 Upon performing Fourier transformation on $h$ and $h^*$, the Fourier components of $h$ and $h^*$ are obtained as follows:
 \begin{eqnarray}
  h_{m}&=&\mathcal{J}_{-m}(\beta)e^{im\theta^{\alpha_{j}}_{\alpha'_{j'}}},\nonumber\\
 (h^{*})_{m}&=&\mathcal{J}_{m}(\beta)e^{im\theta^{\alpha_{j}}_{\alpha'_{j'}}},
 \end{eqnarray}
 where $\mathcal{J}_{-m}(\beta)=(-1)^{m}\mathcal{J}_{m}(\beta)$. Therefore,
$\hat{\mathcal{H}}_{m}$ is obtained in Eq.~(7) as follows:
 \begin{equation}
 \begin{aligned}
 \hat{\mathcal{H}}_{m}&=\sum_{\langle R_j,B_{j'}\rangle}J\left(\mathcal{J}_{-m}(\beta)e^{im\theta^{R_{j}}_{B_{j'}}}\hat{c}^\dag_{R_j}\hat{c}_{B_{j'}}\right.\\
 &\left.+\mathcal{J}_{m}(\beta)e^{im\theta^{R_{j}}_{B_{j'}}}\hat{c}^\dag_{B_{j'}}\hat{c}_{R_{j}}\right)\\
 &+\sum_{\langle G_{j},R_{j'}\rangle}J_1\left(\mathcal{J}_{-m}(\beta)e^{im\theta^{G_{j}}_{R_{j'}}}\hat{c}^\dag_{G_j}\hat{c}_{R_{j'}}\right.\\
 &\left.+\mathcal{J}_{m}(\beta)e^{im\theta^{G_{j}}_{R_{j'}}}\hat{c}^\dag_{R_{j'}}\hat{c}_{G_{j}}\right)\\
 &+\sum_{\langle B_{j},G_{j'}\rangle}J_1\left(\mathcal{J}_{-m}(\beta)e^{im\theta^{B_{j}}_{G_{j'}}}\hat{c}^\dag_{B_j}\hat{c}_{G_{j'}}\right.\\
 &\left.+\mathcal{J}_{m}(\beta)e^{im\theta^{B_{j}}_{G_{j'}}}\hat{c}^\dag_{G_{j'}}\hat{c}_{B_{j}}\right).
 \end{aligned}
 \end{equation}
 \end{appendix}


\begin{references}
\bibitem{su31a} We call
 the case that the Fermi energy falls within the lower bulk gap $1/3$ filling while the case that the Fermi energy falls within
 the upper bulk gap $2/3$ filling.
\bibitem{be65} Other values of the $\gamma_{1}$ and $\gamma_{2}$ are tried with the similar conclusions.
\end{references}

\end{document}